# Modelling the size of Nitrogen c(2x2) islands on Cu(001) with elastic multisite interactions

Wolfgang Kappus

Alumnus of ITP, Philosophenweg 19, D-69120 Heidelberg

wolfgang.kappus@t-online.de

v1.1: 2020-04-05

## ABSTRACT

An extended elastic eigenvector approach for adatom interactions is applied to model Nitrogen c(2x2) atom islands on Cu(001). Oscillating interactions between adatom monomers or dimers are considered. Attractive pair interactions of adatoms are able to describe agglomeration but fail to explain the size of islands. Multisite interactions created by dimers, however, can explain the size of islands if an interaction parameter is adjusted properly. Finite size islands are formed when repulsive monomer interactions are balanced by attractive monomer-dimer interactions.

A simple model of nearest neighbour attraction and $s^{-3}$ repulsion is used as an example for the interaction parameter search.

Previous experimental studies have shown that in the low coverage region Nitrogen atoms agglomerate to isolated square islands of about 5 nm x 5 nm size with a c(2x2) structure. With increasing coverage those islands form regular patterns but do not touch below about 0.4 monolayers. Previous experiments and calculations have shown that the square island pattern show significant elastic effects.

Different adatom structures within the Nitrogen islands have been analyzed, a dimer step model is able to reproduce the experimentally found island size. This model requires a molecular bond between dimers formed by Nitrogen adatoms. The model is also used to compute elastic interactions between islands and to check it against the experimentally found island pattern.

Shortcomings and limitations of the model are discussed and open questions are formulated.

## 1. Introduction

Interactions of adatoms are a subject of continuous interest; various different interaction mechanisms have been described in detail [1]. Adatom structures like superlattices, nanodot arrays, nanostripes, strain relief patterns are interesting for various general and technological reasons; reviews were given in [2,3,4,5]. Lateral interactions also govern the ordering behavior of adatoms and thus the catalytic activities of surfaces [6]. Lattice models are the discrete representations of surface properties and usable for predicting structure and phase diagrams of adatom configurations [1]. Exotic patterns like from N-c(2x2)-Cu(001) islands post a challenge for modelling their behaviour; elastic interactions seem to play a significant role in this system [5].

Strain mediated adatom interactions with the basic $s^{-3}$ distance proportionality and the strong influence of the substrates elastic anisotropy were discussed already in the 1970 decade based on elastic continuum theory [7, 8, 9]. Multisite interactions based on an elastic eigenvector theory have been used to model O adatoms on Pd(001) [10] and Fe adatoms on Cu(111) [11].

In the simple pair interaction model of [8] the interaction of adatoms is mediated by their strain fields generated by single adatoms exerting isotropic stress to their vicinity. Isotropic stress is a consequence of adatom locations on sites with high symmetry. In a lattice description adatoms would exert forces to their immediate substrate neighbours creating a displacement field equivalent to a strain field.

In other cases where adatoms interact directly e.g. via their dipole moment or via electronic overlap, pairs of adatoms may create substrate strain by stretching or compressing the substrate to balance the forces. The strain field of such pairs will mediate interactions between pairs and also between pairs and monomers, in other words trio and quarto multisite interactions. If forces between dimer constituents are central, the model can be kept simple and the resulting stress field can (with symmetry restrictions) be described with one more free parameter [10] than the previous pair interaction model [8]. An oscillating interaction is the consequence of the sharp cutoff of a wave vector integral [10].

Most aspects of adatom multisite interactions have been discussed in detail with a special focus on phase boundary asymmetries [12, 13, 14]. A still unresolved issue in such calculations is the influence longer range pair interactions may have on the determination of trio- and quarto parameters. The issue becomes especially relevant if pair interactions are oscillating in sign.

The N-Cu(001) system, starting from [15] has been studied extensively experimentally and theoretically, a review with references is given in [5].

In short Nitrogen adatoms agglomerate in 5 $\mu$m wide c(2x2) square islands with edges in (100) directions; they form rows at coverage below $\theta \approx 0.1$ ML, a square grid pattern at $\theta \approx 0.25$ ML and come close beyond $\theta \approx 0.35$ ML coverage. Between those islands significant elastic strain was detected hindering coalescence [16]. Beyond $\theta \approx 0.35$ ML coverage (110) directed trenches are observed in completely covered area. The N-Cu(001) system attracted broad interest because it could be used as templates for nanodots [5].

The existence of strain in N-c(2x2) islands and the bare Copper areas between those islands was verified by analysing the electronic structure of Cu(001) [17]. Focus was laid on surfaces states and lattice constant changes were estimated. The strain is suggested to stem from lattice mismatch of the incommensurate c(2x2) Nitrogen lattice on Cu(001) [18]. Boundary types caused by stress relief mechanisms were analyzed in [19].

The question is raised in the following, which interactions are able to create finite islands in the N-Cu(001) system. As one example a nearest neighbour attraction and $s^{-3}$ repulsion model is considered. The other model uses elastic adatom interactions caused by the stress adatoms or dimers exert to the substrate.

The elastic energy of islands with varying width will be calculated and the minimum energy will be searched. Adatom dimers will play a crucial role in this analysis. Inter-island interactions will be derived from the parameters found.

The current semi-heuristic mesoscopic elastic model is seen complementing an exact microscopic ab-initio approach [20].

The paper is organized as follows:
After outlining the general motivation in section 1, a simple model is discussed in section 2 to sketch the method used subsequently. The elastic interaction model is recalled in section 3 and model parameters for N-Cu(001) islands are derived. Those parameters are used in section 4 to derive interactions between islands. In section 5 the models applicability and limitations are discussed and open questions are addressed. Section 6 closes with a summary.

## 2. Nearest neighbour attraction and $s^{-3}$ repulsion model

If adatom nearest neighbour attraction competes with a $s^{-3}$ repulsion within a square island of lattice-gas adatoms, a size with minimal configuration energy can be determined.

The total interaction energy H is assumed to be a sum of pair interactions $V_{ij}$

$$H = \sum_i \sum_{j<i} V_{ij}\, n_i\, n_j \qquad (2.1)$$

with occupation numbers $n_i$ of adatoms $i$ and $j$ within a square island. The pair interactions $V_{ij}$ consist of two parts

$$V_{ij} = V_{ij}^{(1)} + V_{ij}^{(2)} \qquad (2.2)$$

with $V_{ij}^{(1)}=a$ if $i$ and $j$ are nearest neighbours and $V_{ij}^{(1)}=0$ elsewhere; $V_{ij}^{(2)}=bs^{-3}$ where $s$ is the distance between adatoms $i$ and $j$.

With increasing size $w$ (in units of the substrates lattice constant $s_0$) of adatom square islands the total interaction energy $H(w)$ forms a series which can be handled by an analytic power series approximation using the interaction energy per adatom $h(w)=H(w)/w^2$. The power series of $h$ in $w^{-1}$ reads

$$h(w) = \sum_{i=0}^{2} h_i / w^i \qquad (2.3)$$

with $h_i=a*a_i+b*b_i$. Table 1 shows the resulting coefficients $a_i$ and $b_i$

Table 1. Power series coefficients $a_i$ and $b_i$ for the NN / $s^{-3}$ model

| i | $a_i$ | $b_i$ |
|---|---|---|
| 1 | 4 | 8.7167 |
| 2 | -4 | -25.943 |
| 3 | 0 | 35.777 |

Differentiating of the analytic interaction

$$H(w) = \sum_{i=0}^{2} (a*a_i + b*b_i)\, w^{(2-i)} \qquad (2.4)$$

with the coefficients $a_i$ and $b_i$ shown in Table 1 then gives a relation between the interaction strengths $a$, $b$ introduced below Eq. (2.2) and the size $w_{\min}$ where the island energy is minimal. In turn if an island size is given and identified with $w_{\min}$ the ratio $b/a$ can be derived. For $w_{\min}=14$ lattice constants $b/a=0.495$.

The physical nature of the $s^{-3}$ repulsion is left open, a dipole-dipole interaction or a classical elastic interaction could be an example. The strain, however, exerted by adatom dimers is not correctly covered in this model. Therefore an extended elastic model is discussed below.

## 3. Elastic adatom interaction model

In this section the lattice-gas Hamiltonian of Eq. (2.1) is extended to cover sums of discrete multisite interactions as already used to describe elastic interactions of the O-Pd(001) [10] and the Fe-Cu(111) [11] systems.

The total adatom interaction energy $H$ (omitting site energies) is assumed to be a sum of pair interactions $V_{ij}$,

trio interactions $V_{ijk}$ and quarto terms $V_{ijkl}$

$$H = \sum_i \sum_{j<i} V_{ij}\, n_i\, n_j + \sum_i \sum_{j<i} \sum_{k<j<i} V_{ijk}\, n_i\, n_j\, n_k + \sum_i \sum_{j<i} \sum_{k<j<i} \sum_{l<k<j<i} V_{ijkl}\, n_i\, n_j\, n_k\, n_l \qquad (3.1)$$

with occupation numbers $\{n_i, n_j, ..\}$ of adatoms $\{i,,j,,...\}$ within a square island. The discrete interactions will be taken from a substrate strain mediated interaction model as used in [10] and detailed there. Here only the essentials are recalled.

*3.1. Elastic eigenvector adatom interaction model recalled*

Adatoms or dimers exert forces on their substrate neighbours leading to a displacement of substrate atoms to balance those forces. Such displacements will increase or decrease the energy of neighbouring adatoms or dimers. In a continuum description adatoms and dimers exert stress parallel to the surface leading to substrate strain which in turn can lead to an attraction or a repulsion of neighbouring adatoms or dimers. The strength of such strain mediated interaction depends on the stress adatoms or dimers exert on the substrate and on the stiffness of the substrate.

The elastic energy of a substrate with adatoms in a continuous description is given by the sum of two parts, the energy of the distorted substrate and the energy of adatoms exerting tangential forces on the substrate

$$H_{el} = \frac{1}{2} \int_V \epsilon(r)\, c\, \epsilon(r)\, d\,r + \int_S \epsilon(s)\, \pi(s)\, d\,s \,. \qquad (3.2)$$

Here $\epsilon = [\epsilon_{\alpha\beta}]$ denotes the strain tensor field, $c = [c_{\alpha\beta\mu\nu}]$ denotes the elastic constants tensor, and $\pi = [\pi_{\mu\nu}]$ denotes the force dipole- or stress tensor field. The integrals comprise the bulk V or the surface S. The strain field $\epsilon(r)$ is related to the displacement field $u(r)$ by

$$\epsilon_{\alpha\beta}(r) = \frac{1}{2}(\nabla_\alpha u_\beta(r) + \nabla_\beta u_\alpha(r))\,. \qquad (3.3)$$

Following [8] the stress field $\pi(s)$ is superimposed of $n_t$ different types $\pi_k(s)$

$$\pi(s) = \sum_{k=1}^{n_t} \pi_k(s) = \sum_{k=1}^{n_t} P_k\, \rho_k(s)\,, \qquad (3.4)$$

where we have introduced $n_t$ different types of stress tensors $P_k$ and of adatom monomer or dimer densities $\rho_k(s)$.

On (001) surfaces with adatom positions on high symmetry sites $n_t=3$ i.e. $k$ can take the values 1 to 3 where $P_1 = P_1[\delta_{\alpha\beta}]$ stand for an isotropic monomer stress tensor (or force dipole tensor),

$P_2 = P_2[\delta_{\alpha 1}\delta_{\beta 1}]$ stands for an anisotropic dimer stress tensor. $P_3$ is the 90° rotated equivalent of $P_2$. On (001) surfaces with adatom positions on high symmetric sites $P_3 = P_2$.

$\rho_1(s)$ then stands for the adatom monomer density distribution, $\rho_2(s)$ for the x-directed adatom dimer density distribution and $\rho_3(s)$ for the rotated equivalent of $\rho_2(s)$.

With Eq. (3.4) Eq. (3.2) reads

$$H_{el} = \frac{1}{2} \int_V \epsilon(r)\, c\, \epsilon(r)\, d\,r + \sum_{k=1}^{n_t} \int_S \epsilon(s)\, P_k\, \rho_k(s)\, d\,s \,. \qquad (3.5)$$

The strain field $\epsilon(r)$ is determined for given densities $\rho_k(s)$ by the requirement of mechanical equilibrium

$$\delta H_{el}/\delta u_\alpha(r) = 0\,. \qquad (3.6)$$

After an expansion in plane waves and calculations beyond the scope of this paper we end up at [10]

$$H_{el} = \frac{1}{2} \sum_{k=1}^{n_t} \sum_{l=1}^{n_t} \int_S \int_S \rho_k(s)\, V_{kl}(s-s')\, \rho_l(s')\, d\,s\, d\,s'\,, \qquad (3.6)$$

with the elastic interaction $V_{kl}$ between adatoms (or dimers) of type $k$ at $0$ and adatoms (or dimers) of type $l$ at

$$\mathbf{s} = (s, \chi)$$

$$V_{kl}(\mathbf{s} - \mathbf{s'}) = (2\pi)^{-1} \sum_p \omega_{kl,p} \cos(p\,\chi) \cos(p\,\pi/2) \int_0^{\kappa_{BZ}} \kappa^2 J_p(\kappa s)\, d\kappa, \qquad (3.7)$$

where $J_p(\kappa s)$ denotes the Bessel function of order $p$ and $\omega_{kl,p}$ denotes eigenvalues to be discussed below. The $\kappa$ integral with a hard limit at the surface Brillouin zone $\kappa_{BZ}$ results in a generalized hypergeometric function which together with its coefficients decays oscillating like $s^{-3/2}$. In the sections below the interaction

$$V_{kl}(s,\chi) = (2\pi)^{-1} \sum_p \omega_{kl,p} \cos(p\,\chi) \cos(p\pi/2)\, 2^{-1-p}\, \kappa_{BZ}{}^3\, (s\,\kappa_{BZ})^p\, \Gamma\!\left(\frac{3+p}{2}\right) *$$

$$_1F_2\!\left(\left(\frac{3+p}{2}\right),\left(\frac{5+p}{2}, 1+p\right), -\frac{1}{4}s^2\,\kappa_{BZ}{}^2\right) X(s), \qquad (3.8)$$

with the heuristic factor

$$X(s) = (s/s_0)^{-3/2} \qquad (3.9)$$

is used, ensuring proper mesoscale $s^{-3}$ decay and a regular behaviour for $s \geq s_0$; $s_0$ is the lattice constant of Cu. In (3.9) $\Gamma(p)$ denotes the Gamma function and $_1F_2(a,b_1,b_2,s)$ the generalized hypergeometric function. $_1F_2$ is the appropriate name for the $_pF_q$ function with one factor $a$ and two factors $b_i$ in Eq. (3.8). $\kappa_{BZ}$ is the wave vector distance between the origin and the surface Brillouin zone in $\chi$ direction. $\omega_{kl,p}$ are eigenvalues for the 4x4 interaction types $(k,l)$ with $p$ resulting from a series expansion. $p$ takes the values 0, 4, 8 in the monomer case and 0, 2, 4 in the dimer case. The eigenvalues $\omega_{kl,p}$ are proportional to the product of stress parameters $P_k P_l$ and inversely proportional to the elastic constant $c_{44}$ defining a dimensionless constant $\hat{\omega}_{kl,p}$ by

$$\omega_{kl,p} = \hat{\omega}_{kl,p}\, P_k P_l / c_{44}. \qquad (3.10)$$

Tables 2 and 3 show the dimensionless coefficients $\hat{\omega}_{kl,p}$ for Cu(001) using elastic constants $c_{11}$=169, $c_{12}$=122, $c_{44}$=75.3 GPa [21].

The $k=l=1$ coefficients belong to a monomer-monomer (pair) interaction $V_{ij}$ in Eq. (3.1).

The $k=1$, $l=2$ coefficients belong to monomer-dimer (trio) interactions $V_{ijk}$ in Eq. (3.1).

The $k=2$, $l=2$ coefficients belong to (quarto) interactions of two parallel dimers,

The $k=2$, $l=3$ coefficients belong to (quarto) interactions of two dimers spanning an angle of 90°, both $V_{ijkl}$ in Eq. (3.1).

$X(s)$ in Eq. (3.9) slightly differs from a similar heuristic factor in Eq. (2.14) of [10].

Table 2. Pair interaction coefficients $\hat{\omega}_{11,p}$ for Cu(001) calculated with the elastic eigenvector approach of [10].

| k | l | $\hat{\omega}_{kl,0}$ | $\hat{\omega}_{kl,4}$ | $\hat{\omega}_{kl,8}$ |
|---|---|---|---|---|
| 1 | 1 | −1.0538 | −0.1363 | 0.0018 |

Table 3. Trio and quarto interaction coefficients $\hat{\omega}_{kl,p}$ for Cu(001) calculated with the elastic eigenvector approach of [10].

| k | l | $\hat{\omega}_{kl,0}$ | $\hat{\omega}_{kl,2}$ | $\hat{\omega}_{kl,4}$ |
|---|---|---|---|---|
| 1 | 2 | −0.7439 | −0.84018 | −0.096410 |
| 2 | 2 | −1.2705 | −1.1882 | 0.082265 |
| 2 | 3 | 0.218550 | 0 | −0.218650 |

3.2. Surface Brillouin zone

Due to the oscillating interaction in Eq. (3.8) the choice of the surface Brillouin zone shape has a significant influence. Like in [10] a square approximation

$$\kappa_{BZ}(\chi) = \kappa_{BZ0}(15/16 + \cos(4\chi)/16) \ . \tag{3.11}$$

was selected, ensuring $\kappa_{BZ}(0) = \kappa_{BZ0} = fbz*2\pi$. This choice reflects the symmetry on the (001) surface, the value 16 is chosen to give zero curvature in the (100) direction and the factor *fbz*≈1 will be detailed in section 4.

### 3.3. Fitting the lattice sum by a power series for deriving stress parameters

Following [10] the elastic energy will now be formulated as lattice sums for 3 parts, consisting of a monomer-monomer part, a monomer-dimer part, and a dimer-dimer part, which allows calculating the elastic energy of an adatom island as function of the stress parameters $P_1$ and $P_2$.

$$H_{el} = P_1 P_1 \hat{H}_{11} + P_1 P_2 \hat{H}_{12} + P_2 P_2 \hat{H}_{22}. \tag{3.12}$$

where the $\hat{H}_{ik}$ are partial lattice sums independent of $P_1$ and $P_2$. $\hat{H}_{ik}$ replace the $h_{ik}$ used in [10] to avoid confusion with the energies per adatom as used in Eq. (2.3).

Eq. (3.12) now allows a procedure as used in section 2 to calculate numerically the $\hat{H}_{ik}(w)$ for a series of square islands of size $w$, to fit the series

The method used in section 2 will subsequently be adapted: For a series of square island sizes $w$ partial elastic energies $\hat{H}_{ik}(w)$ will be calculated and fitted by an analytical expression. The partial elastic energies per adatom

$$\hat{h}_{ik}(w) = \hat{H}_{ik}(w)/w^2 \tag{3.13}$$

are best suited to be fitted by a power series in $w^{-1}$

$$\hat{h}_{ik}(w) = \sum_{m=0}^{2} \hat{h}_{ik}^{(m)}/w^m + O(w^3) . \tag{3.14}$$

The $\hat{h}_{ik}^{(m)}$ will subsequently be derived using 3 nodes of calculated $\hat{h}_{ik}(w)$ values. The analytic energy per island adatom is with Eqs. (3.12), (3.13), (3.14)

$$h(w) = \sum_{m=0}^{2} \left( P_1 P_1 \hat{h}_{11}^{(m)} + P_1 P_2 \hat{h}_{12}^{(m)} + P_2 P_2 \hat{h}_{22}^{(m)} \right)/w^m . \tag{3.15}$$

To width of an island is assumed to be given by the minimum its energy

$$H(w) = \frac{1}{2} \sum_{m=0}^{2} \left( P_1 P_1 \hat{h}_{11}^{(m)} + P_1 P_2 \hat{h}_{12}^{(m)} + P_2 P_2 \hat{h}_{22}^{(m)} \right) * w^{(2-m)} . \tag{3.16}$$

So the minimum width $w_{min}$ can be taken from

$$0 = \sum_{m=0}^{1} 2^{(1-m)} \left( P_1 P_1 \hat{h}_{11}^{(m)} + P_1 P_2 \hat{h}_{12}^{(m)} + P_2 P_2 \hat{h}_{22}^{(m)} \right) * w_{min}^{(1-m)} . \tag{3.17}$$

If otherwise the width $w_{min}$ is given (e.g. by experiments), the stress parameter $\hat{P}_2 = P_2/P_1$ is a solution of the quadratic equation

$$\left(2*w_{min}* \hat{h}_{22}^{(0)} + \hat{h}_{22}^{(1)}\right) \hat{P}_2^2 + \left(2*w_{min}* \hat{h}_{12}^{(0)} + \hat{h}_{12}^{(1)}\right) \hat{P}_2 + \left(2*w_{min}* \hat{h}_{11}^{(0)} + \hat{h}_{11}^{(1)}\right) = 0 \ . \tag{3.18}$$

With the solution of Eq. (3.18) a stress parameter $\hat{P}_2$ is found which minimizes the elastic energy Eq. (3.12) for a given island width $w_{min}$.

### 3.4. Nitrogen c(2x2) configurations within square island

N-c(2x2) configurations within square islands are not unique:
If Nitrogen adatoms interact equally with all 4 nearest neighbours a homogeneous island arises.

If, however, Nitrogen adatoms form dimers (e.g. with some molecular bond) a variety of internal structures could arise, like chess pattern or step pattern.

Examples are shown in Fig.1.a to c.

The black circles stand for Cu atoms at the (001) surface, the red discs indicate c(2x2) Nitrogen adatoms above 4 fold coordinated hollow sites, in certain cases slightly displaced. Overlapping disks indicate interacting adatoms. Size and displacement of disks are oversubscribed for better distinction.

Fig.1.a depicts a homogeneous placement of Nitrogen adatoms just above 4 fold coordinated hollow sites.

Fig.1.b. depicts a chess type order of Nitrogen dimers. The Nitrogen dimer constituents are assumed closer due an attractive interaction.

Fig.1.c. depicts a step order of Nitrogen dimers. Again Nitrogen dimer constituents are assumed closer due an attractive interaction.

All 3 order types are increasingly isotropic with increasing island size. Uneven sized chess pattern and all step order patterns contain monomers at edges or corners since some adatoms lack a partner. The ratio of monomers to dimers decreases with increasing island size.

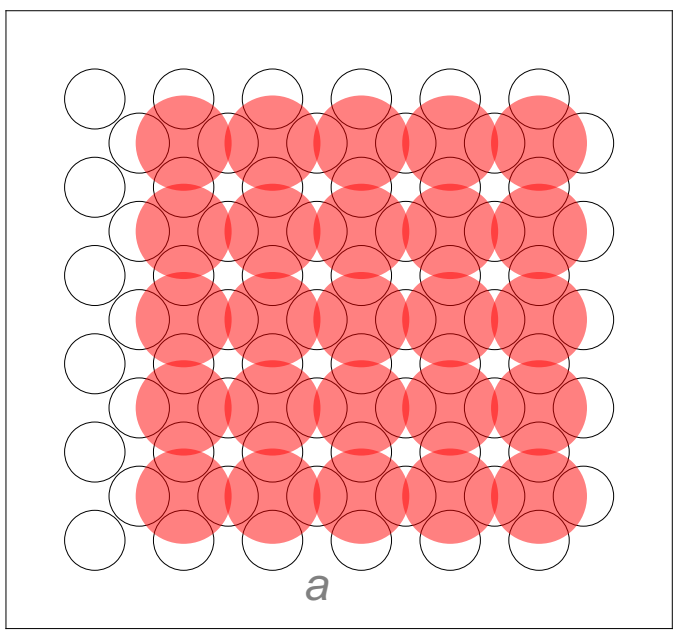

*a*

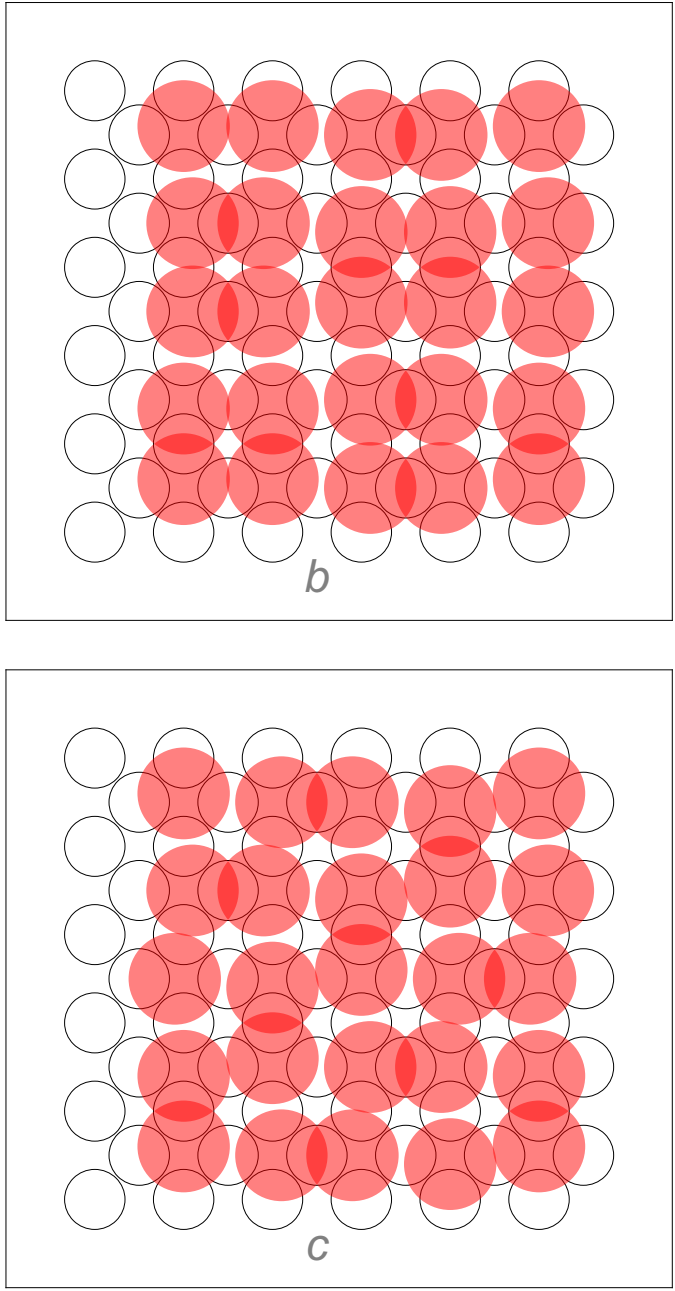

Fig. 1. Different Nitrogen c(2x2) adatom or dimer configurations within an island of 5 lattice constants. Black circles indicate Cu substrate atoms, red disks indicate N-adatoms; dimers show overlap indicating electronic interaction.
a. homogeneous monomers placement,
b. chess type dimer placement,
c. step order dimer placement.

### 3.5. Partial lattice sums

Unfortunately the elastic interaction Eq. (3.8) with (3.9) contains heuristic approximations, so any result and conclusion carries uncertainties. The method to derive elastic parameters, however, seems to be valid also for modifications of Eq. (3.8) and (3.9).

The experimentally found N-c(2x2) island width of 5 $\mu$m [5] or 14 Cu $s_0$ lattice constants seems to be a solid starting point for the analysis.

For 3 different placements shown in Fig.1 and for different surface Brillouin zones the partial lattice sums $\hat{H}_{ik}(w)$ of Eq. (3.12) and subsequently the $\hat{h}_{ik}^{(m)}$ values of Eq. (3.13) have been calculated with the interaction $V_{kl}(s, \chi)$ of Eqs. (3.8) and (3.9). All but one have failed to produce real solutions of Eq. (3.18), so the remaining step placement of Fig.1.c and the surface Brillouin zone of Eq. (3.11) is the base for the results presented below.

The aforementioned partial lattice sums per adatom $\hat{h}_{ik}(w)$ of Eqs. (3.12) and (3.13) were calculated numerically for island widths up to 24 $s_0$. They are shown in Fig. 2 as dots together with their fitted counterparts as curves. The fits are appropriate for islands greater than 4 lattice constants. The pair interaction component $\hat{h}_{11}(w)$ is repulsive, the multisite components $\hat{h}_{12}(w)$ and $\hat{h}_{22}(w)$ are attractive for larger islands.

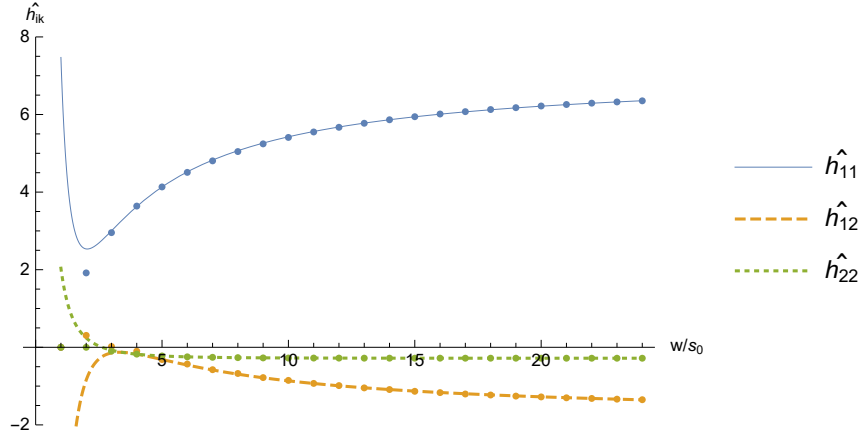

Fig. 2. Partial lattice sums per adatom $\hat{h}_{ik}(w)$ for island widths up to 24 lattice constants (as dots). $\hat{h}_{11}(w)$ represents the pair interaction, $\hat{h}_{12}(w)$ the trio interaction and $\hat{h}_{22}(w)$ the dimer-dimer interaction component in Eqs. (3.12) and (3.13). The associated fitted curves are derived from the power series Eq. (3.14).

The $\hat{h}_{ik}^{(m)}$ values for fitting the $\hat{h}_{ik}(w)$ components with power series according to Eq. (3.14) are shown in Table 4. The $\hat{h}_{ik}^{(0)}$ values represent partial energy per adatom in island of infinite size; the other values stand for the slope of the $\hat{h}_{ik}(w)$ for fitting smaller islands.

Table 4. $\hat{h}_{ik}^{(m)}$ values for fitting the $\hat{H}_{ik}(w)$ components with power series according to Eq. (3.14)

| m | $\hat{h}_{11}^{(m)}$ | $\hat{h}_{12}^{(m)}$ | $\hat{h}_{22}^{(m)}$ |
|---|---|---|---|
| 0 | 7.09802 | -1.77871 | -0.275374 |
| 1 | -18.6232 | 10.957 | -0.251061 |
| 2 | 18.996 | -18.1785 | 2.56428 |

*3.6. Minimal island energy*

With the $\hat{h}_{ik}^{(m)}$ values of Table 4 it is now straightforward to solve Eq. (3.18) for the stress parameter $\hat{P}_2$= -7.7853 corresponding to a minimum analytic elastic energy $H(w)$ at $w_{\min}$. The negative sign of $P_2$ means contractive stress dimers exert to the substrate. $H(w)$ according to Eq. (3.16) is plotted as curve in Fig. 3 together with the discrete energy values of 24 different island sizes. $H(w)$ is given in $P_1^2$ units since independent of $P_1^2$. Again the fit is reasonable.

Small changes of $\hat{P}_2$ would alter the slope of $H(w)$ significantly: decreasing $\hat{P}_2$ by 2% would shift its minimum $w_{\min}$ from 14 to 20 lattice constants.

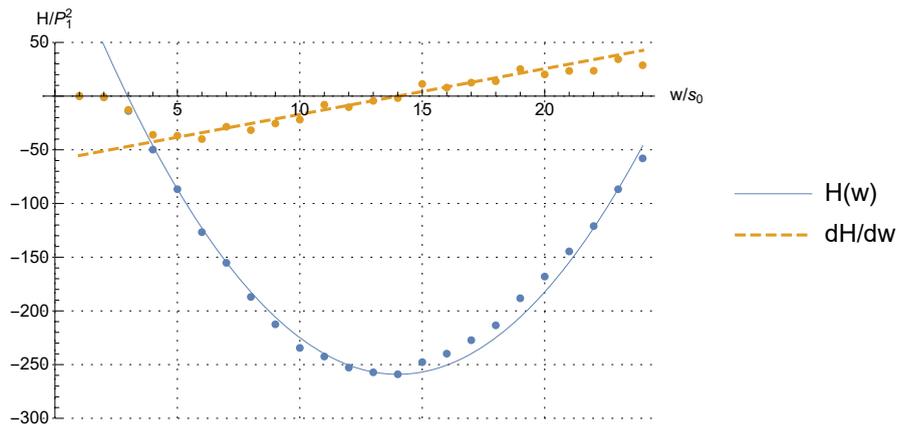

Fig. 3. Analytic elastic energy $H(w)$ according to Eq. (3.16) as curve together with the discrete energy values of 24 varying island sizes. The derivative $dH(w)/dw$ together with the energy differences are plotted too.

The broad $H(w)$ energy minimum around $w_{min}=14$ lattice constants would in thermal equilibrium lead to a size- and shape distribution of Nitrogen islands. Experiments show a distribution of more or less square island around 14 Cu lattice constants [5]. Kinetic growth is also considered. Kinetic growth stops when additional adatoms would increase the island energy; the $dH(w)/dw$ line indicates this moment in Fig.3 when crossing the abscissa. The jumps of the energy differences around the $dH(w)/dw$ line are explained by varying amounts of monomers within the Nitrogen islands.

If a monomer is placed on a Nitrogen island corner it lacks most of the attractive trio interaction, so it may leave the island and a void is left. This could explain trimmed corners in experiments [5].

### 3.7. LEED pattern simulation

Since the c(2x2) structure of Nitrogen on Cu(001) is based on LEED experiments and broadly accepted, the difference between LEED patterns of an undistorted c(2x2) structure of Fig.1.a and the step structure shown in Fig.1.c. needs to be analyzed. For this purpose a simulation was conducted by calculating the Fourier Transform of 300x300 occupation matrices. The undistorted structure had an "1" every 20 horizontal or transversal positions; in the step structure of Fig.1.c adatoms had deviations of one horizontal or vertical position from the undistorted structure. This simulates a constricting strain of 10% due to adatom dimers; the 10% value is chosen for demonstration only.

While the Fourier Transform of the undistorted c(2x2) structure shows the regular *(2i,2k)* pattern, the Fourier Transform of the distorted structure in Fig. 4 shows 45° rotated square domains of regular spot patterns separated by unstructured 45° directed stripes. Lines of spots show smeared steps of 1/4 the spot distance. The pattern can roughly be interpolated by *(2.05i,2.05k)*.

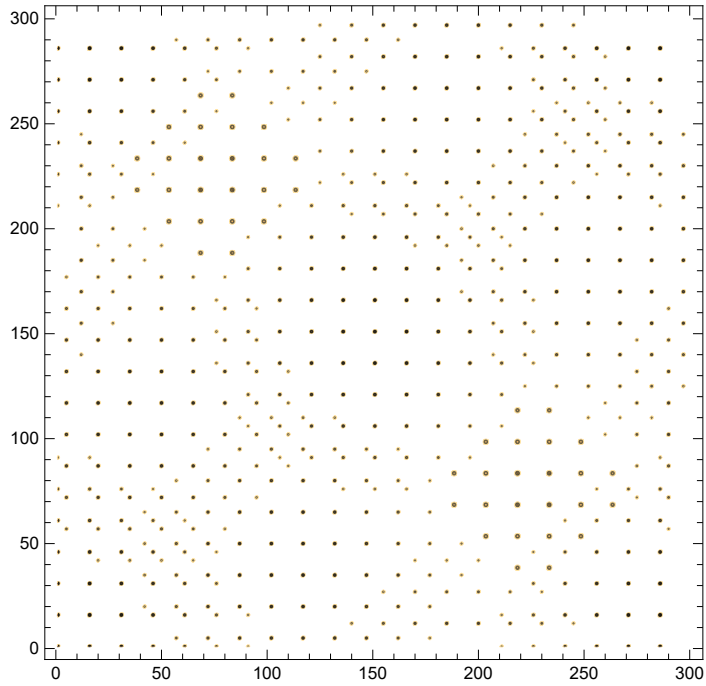

Fig. 4. LEED pattern of the N-Cu(001) structure of Fig.1.c simulated, assuming 10% constricting dimer strain

The key message of this simulation is that inspecting the low index domain of about $i+j<4$ is not enough to decide upon the N-Cu(001) adatom structures of Fig.1.a and c from LEED experiments.

The difference of LEED patterns of an undistorted c(2x2) structure and those of a structure expanded by isotropic stress is not discussed because the results of the previous section indicate some balance between the expansive isotropic stress and the constricting stress of dimers.

---

# 4. Elastic interactions between square islands

The grid like-pattern of square islands observed for the N-c(2x2)-Cu system at coverages between about 0.1 and 0.4 ML [5] raises the question if the elastic interaction with the parameters of section 3 is able to comply with the observations.

The elastic energy $E$ of square island pairs of width $w_{min}$ separated by a distance $d$ in (100) direction was calculated. The difference of $E$ and twice the elastic energy $H(w_{min})$ is the island interaction energy $H_{i(100)}(d)$. Likewise the island interaction energy $H_{i(110)}(d)$ of islands situated in (110) direction is calculated; here $d$ means the distance to the projection in (100) direction, so $d$ means the channel width of an island grid.

In Table 5 the calculated inter island interaction energies in $P_1^2$ units are listed for 12 different distances in Cu lattice constant units are listed. For convenience the coverages $\theta(d)$ of perfect island grids with channel widths $d$ are also shown starting with c(2x2) half Nitrogen coverage at zero width $d$.

Table 5. Inter-island interaction energies $H_{i(100)}(d)$ and $H_{i(110)}(d)$ as function of the channel width $d$ in $P_1^2$ units calculated with the parameters of section 3. The coverage $\theta$ is given in Nitrogen monolayers units.

| $d / s_0$ | 0 | 1 | 2 | 3 | 4 | 5 | 6 | 7 | 8 | 9 | 10 | 11 |
|---|---|---|---|---|---|---|---|---|---|---|---|---|
| $\theta(d)$ ML | 0.5 | 0.44 | 0.38 | 0.34 | 0.30 | 0.27 | 0.24 | 0.22 | 0.20 | 0.18 | 0.17 | 0.16 |
| $H_{i(100)}(d) / P_1^2$ | 378 | 362 | 233 | 121 | 55 | 23 | -7.5 | -21 | -33 | -31 | -39 | -34 |
| $H_{i(110)}(d) / P_1^2$ | 27 | 13 | 5 | 1.0 | 0.2 | 0.5 | 0.7 | 0.4 | 0.2 | 0.04 | 0.08 | 0.1 |

A strong anisotropy is observed; in (100) direction the interaction is repulsive up to $d=5$ $s_0$ and attractive beyond, in (110) direction the interaction is weakly repulsive and almost neglectable. Such interaction is in line with a dense packing of square islands as observed in a coverage range 0.25 ML $\lesssim \theta \lesssim$ 0.5 ML. In the range below the attractive (100) interaction would arrange for local grids of $\theta \approx 0.25$ ML. In thermal equilibrium the observed widths may of course deviate from the theoretical values; for example a continuous layer could coexist with a grid pattern in the neighbourhood for $\theta \approx 0.4$ ML. Another deviation from a perfect square island grid pattern with distances $d$ in (100) and (010) direction could be a *(d+1)\*(d-1)* pattern with almost the same coverage and interaction energy.

The existence of (100) directed chains instead of a square island pattern in the low coverage range is explained by the attractive inter-island interaction in the $d \geq 6 \, s_0$ region.

The guiding line $H_{i(100)}(d) / P_1^2 \approx a_1 + a_2/d$ in Fig.5 shows a strong $d^{-1}$ interaction. The dots indicate a regular behavior when islands touch at $d \to 0$.

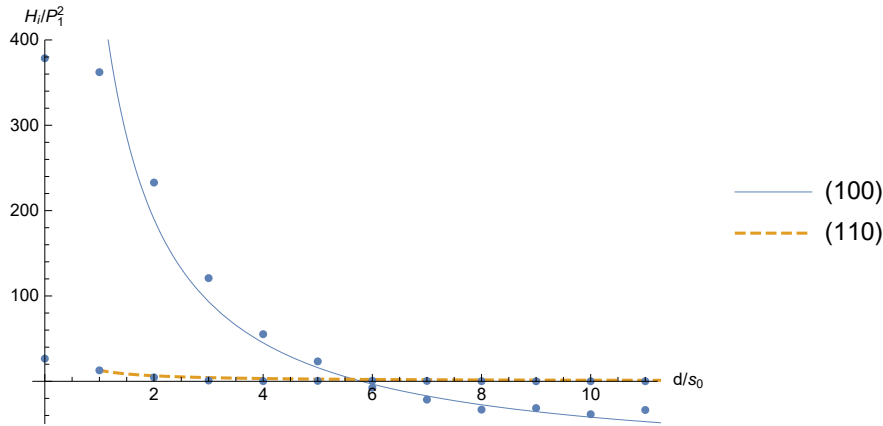

Fig 5. Inter-island interaction energies $H_{i(100)}(d)$ and $H_{i(110)}(d)$ as function of the channel width $d$ in $P_1^2$ units calculated with the parameters of section 3.6.

Coming back to the choice of $\kappa_{BZ0}$ and *fbz* respective in Eq. (3.11) which significantly influences both the value of $\hat{P}_2$ and thus the inter-island interactions in Table 5, it is noted that the currently used *fbz*=0.97 leads to an attraction of $H_{i(100)}(d) / P_1^2$ for $d \geq 6 \, s_0$ while the value *fbz*=1 does not. A chess pattern at coverage $\theta \approx$ 0.25 ML is therefore not expected with the choice *fbz*=0.97.

# 5. Discussion

Two adatom interaction models have been considered for the finite size of N-c(2x2) island. A simple one where nearest neighbour attraction competes with a $s^{-3}$ repulsion, and a less simple one with 3 parts of an elastic interaction. For both parameters can be found with energy minima at 14 lattice constants. Both models contain heuristic elements, the first one lacks consistency while missing stress effects to the Cu substrate, the second seems to be more consistent implicitly considering the electronic interactions between adjacent Nitrogen adatoms. This approach seems valid if this electronic interaction is compensated by the elastic energy of a dimer.

Both models are far from the precision of an ab-initio approach which calculates interactions on an atomic scale. Since the elastic approach covers a mesoscopic range there could be a fruitful coo petition. The discussion below concentrates on the elastic model because is seems to be more realistic.

### 5.1. Restrictions and limitations of the elastic model

The elastic eigenvector model is based on the theory of elasticity in the substrate and on the lateral stress adatoms apply to the surface. Key assumption is the mechanism by which adatoms and dimers interact. Monomer adatoms sitting on high symmetry adatom sites expand or contract the substrate by creating isotropic stress. Multisite interactions are due to the anisotropic stress adatom pairs exert to the substrate when they are bound electronically and stretched (or compressed) due to their position on substrate sites. Multisite effects [12] are essential for the elastic model in the N-c(2x2)-Cu(001) case.

The existence of N-N dimers within the c(2x2)-N island is key for the model. Currently there is a lack of obvious arguments from experiments and theory. The existence of optical surface phonon modes without significant dispersion within Nitrogen islands [22], however, seems to support the existence of N-N bonds; optical phonon modes without dispersion are characteristic for local oscillations of higher frequency. The LEED pattern simulation of section 3.7 is intended to provide some more credibility to the current basic assumption.

The choice of the step order dimer placement in Fig. 1.c. may appear strange, but arguments like 90° symmetry and the existence of (110) trenches at coverages beyond 0.35 ML [5] support its likeliness.

At very small distances electronic interactions dominate, typically decaying with $s^{-5}$; at larger distances surface-state electronic interactions dominate decaying with $s^{-2}$ [23]. In the intermediate mesoscopic range elastic interactions prevail at least for the Fe-Cu(111) system [10] and the O-Pd(001) system [11]. The mesoscale restriction is also valid for the present continuous elastic model; it cannot be expected to cover very small distances. An elastic continuum model for the substrate describes well mesoscopic range effects and elastic anisotropies (strong in case of adatom dimers).

The elastic eigenvector model pioneered in [24] and used in this analysis is part of a family of elastic methods [4, 25, 26] used to describe elastic effects on surfaces.

Further shortcomings are related with the surface Brillouin zone shape and the $\kappa$ integration cutoff. The shape is not the natural choice and the sharp cutoff leads to a function which does not follow the classical $s^{-3}$ law though it is oscillating. A heuristic correction is necessary; modifications of this would, of course, also slightly modify the results.

The restriction to high symmetry adatom locations has the advantage of stress parameter $P_k$ degeneracy. On (001) surfaces $P_3=P_2$ due to the equivalence of dimers with a 90° angle. This reduces the number of free model parameters.

The concentration to square islands without trimmed corners seems reasonable and does not compromise the overall results.

A further key assumption is an ideal flat surface, i.e. the absence of steps.

### 5.2. Open questions and further aspects

The search for a cutoff function in Eq. (3.7) between a smooth exponential $Exp(-\kappa^2 s^2)$ and a hard Heaviside function $\theta(\kappa\text{-s})$ could lead to an improved oscillating interaction with a proper decay and phase. As shown in [8] the smooth exponential cutoff leads to elastic interactions with an $s^{-3}$ decay without oscillation.

DFT calculations of short distant adatoms and dimers should improve the model base. Analysing the range of possible monomer to dimer- and dimer to dimer placements, in various sites and orientations, would add insight into the nature of the multisite interaction. Such calculations would also serve as tests for the elastic interactions used.

A theoretical (DFT) model to determine the magnitude of the stress parameters $P_k$ could determine the elastic adatom interaction ab initio. For this purpose the lateral displacements of adjacent substrate atoms an adatom creates would have to be calculated as well as the lateral forces between adatom and those substrate atoms.

Further experiments and calculations to qualify the dimer assumption and their step order placement choice would be interesting.

# 6. Summary

The previously experimentally found size of Nitrogen islands on Cu(001) has been modelled by simple short range and less simple elastic interactions. Lattice sums of variable island size have been fitted by analytic functions allowing to determine model parameters for which the energy in minimal. The size of islands with minimal energy was assumed to equal the experimental value. The elastic interaction consists of pair- and multisite parts. The energy minima are broad; in thermal equilibrium island sizes would show a certain distribution. Interactions between islands have been derived using the infra island parameters. Limitations of the model are analyzed and open questions are addressed.

# Acknowledgements

This work is dedicated to my wife thanking for her continuous caring support.